\documentclass[12pt]{article}

\begin{document}
\title{ {\bf ${\cal N}=2$ supergravity in three dimensions and its G\"odel
supersymmetric background}}
\author{M. Ba\~nados$^{(a)}$, A. T. Faraggi$^{(b)}$ and S. Theisen$^{(c)}$ \\
 $^{(a)}$Departamento de F\'{\i}sica,  Pontificia Universidad Cat\'{o}lica de Chile,
Casilla 306, \\ Santiago 22, Chile \\
$^{(b)}$ Department of Physics, University of Michigan,\\
Ann Arbor, MI 48109-1120, USA\\
  $^{(c)}$Max-Planck-Institut f\"ur Gravitationsphysik,
Albert-Einstein-Institut,\\
 14476 Golm, Germany \\
 {\tt maxbanados@fis.puc.cl, faraggi@umich.edu, theisen@aei.mpg.de}
 }
\date{}

\maketitle

\begin{abstract}

The four dimensional G\"odel spacetime is known to have the structure
$M_3\times \Re$.  It is also known that the three-dimensional factor
$M_3$ is an exact solution
of three-dimensional gravity coupled to a Maxwell-Chern-Simons
theory.  We build in this paper a ${\cal N}=2$ supergravity extension for
this action and prove that the G\"odel background preserves half of all
supersymmetries.

\end{abstract}

G\"odel-type solutions to general relativity have recently been
under scrutiny due to the  discovery  \cite{Gauntlett-} of their
supersymmetric properties. Black holes on these backgrounds have
also been found \cite{Gimon-}. Since these black holes have unusual
asymptotics, issues like first law of thermodynamics and the proper
definition of charges are subtle and require detailed analysis (see
\cite{Barnich-} and references therein for a detailed discussion).
G\"odel spacetimes in string theory have been considered in
\cite{Harmark-} and \cite{Israel:2003cx}.

Most of the discussion that followed the work \cite{Gauntlett-} concerned
a class of G\"odel solutions existing in five dimensions. As it is well-known
this theory contains in its bosonic sector the gravi-photon, that is the
metric coupled to a Maxwell theory with the addition of a Chern-Simons
term $AdAdA$.

On the other hand, the original four dimensional G\"odel spacetime,
discovered in 1949, has a direct product structure $M_4=M_3\times\Re$
where the three dimensional factor $M_3$ encodes most of its interesting
properties.  Motivated by \cite{Gauntlett-} it was shown in \cite{BBCG}
that indeed the $M_3$ factor is a solution to three-dimensional gravity
coupled to a Maxwell theory including the 3d Chern-Simons term $AdA$.
Particles and black holes on this background were also discussed
in \cite{BBCG}.  The next step, which we take in this paper, is to study
the supersymmetric properties of this solution.

Consider the bosonic action,
\begin{eqnarray}
I[g_{\mu\nu},A_\mu] = \frac{1}{16\pi G} \int d^3x \left[\sqrt{-g}\left(R
+\frac{2}{l^2} - \frac{1}{4} F_{\mu\nu}F^{\mu\nu}\right)
-\alpha\,\epsilon^{\mu\nu\rho}A_\mu
\partial_{\nu} A_\rho\right]. \label{action0}
\end{eqnarray}
The field,
\begin{eqnarray}\label{Godel}
ds^2 &=& -dt^2 - 4\alpha r dt d\varphi +
\left[2r-\left(\alpha^2l^2 -1 \right){2r^2 \over l^2} \right]d\varphi^2  +
\left(2r+(\alpha^2l^2+1){2r^2 \over l^2} \right)^{-1} dr^2\,   \label{Godel2}\\
A &=& \sqrt{\alpha^2l^2 -1}\, {2r \over l}\, d\varphi\,. \label{gauge}
\end{eqnarray}
is an exact solution of the equations deriving from (\ref{action0})
\cite{BBCG}. The metric represents the 3d factor $M_3$ of the original
G\"odel solution (actually, its generalization containing two parameters,
$l$ and $\alpha$ \cite{Reboucas-}). Given the high symmetry of this
solution -- it has four Killing vectors --
it is a natural question to ask whether it preserves some
supersymmetries.

Note the strong similarities between the 3d bosonic action (\ref{action0})
and the corresponding 5d supergravity action. Nonetheless, the supergravity
theory corresponding to (\ref{action0}) has some subtleties.  In particular
we would like to have the cosmological radius $l$ and Chern-Simons coupling
$\alpha$ as arbitrary parameters.

The minimal ${\cal N}=1$ supergravity extension to (\ref{action0})
consists of two super-multiplets:\footnote{We work on-shell, i.e. no
auxiliary fields appear.} the gravity multiplet
$\{g_{\mu\nu},\psi_\mu\}$ and a vector multiplet
$\{A_\mu,\lambda\}$. $\psi_\mu$ is the spin 3/2 Rarita-Schwinger
field, while $\lambda$ is a spin 1/2 fermion. Both are Majorana
fermions. However, it is easy to see (assuming an action with no
higher derivatives) that the G\"odel background cannot be a
supersymmetric solution to this system. In fact, the transformation
for the Majorana spinor field $\lambda$ has the form $\delta \lambda
= F^{\mu\nu}\gamma_{\mu\nu}\epsilon.$ For the background
(\ref{gauge}) $F^{\mu\nu}$ is non-zero and one easily verifies that
the equation $\delta \lambda=0$ implies $\gamma_0 \epsilon=0$ and
hence $\epsilon=0$.

We thus explore extended supergravity, or, more precisely,
the three-dimensional ${\cal N}=2$ vector supermultiplet
coupled to ${\cal N}=2$ supergravity.
In three dimensions the former consists of a vector, a real scalar
and complex Dirac fermion, $\{A_\mu,\phi,\lambda\}$.  The gravity
multiplet contains \cite{AT,Witten88} the metric
$g_{\mu\nu}$, a complex Rarita-Schwinger field
$\psi_\mu$, and an Abelian $U(1)$ gauge field $B_\mu$.  The field $B_\mu$ is
independent of $A_\mu$. One might think that the gravity multiplet
suffices for our purposes. However, at the level of at most two derivatives,
the gauge field only enters though the Chern-Simons term and it does not
allow for the free parameter $\alpha$.

The full ${\cal N}=2$ Lagrangian (to quadratic order in the fermions)
is \footnote{We have used the following conventions:
Our metric $g_{\mu\nu}$ has signature $(-,+,+)$. The Dirac matrices satisfy
$\lbrace\gamma^\mu,\gamma^\nu\rbrace=2 g^{\mu\nu}$ and
$\gamma^\mu\gamma^\nu=g^{\mu\nu}+\epsilon^{\mu\nu\rho}\gamma_\rho$.
$\epsilon^{\mu\nu\rho}=\pm1,\,\epsilon_{\mu\nu\rho}=\mp1$.
$\bar\lambda=\lambda^\dagger\gamma_0$.
$(\gamma^\mu)^\dagger=\gamma_0\gamma^\mu\gamma_0$.
In the 1.5 formalism the variation of the spin connection is obtained from
its algebraic equation of motion, and not needed explicitly. }
\begin{equation}\label{action}
L ={1 \over \kappa^2} L_{-2} + L_0 + \kappa L_1 +\kappa^2 L_2
\end{equation}
where
\begin{eqnarray}
L_{-2} &=& {1 \over 2} e\left(R+ {2 \over l_0^2} \right)\nonumber\\
L_0 &=& -i\epsilon^{\mu\nu\rho} \bar\psi_\mu D_\nu \psi_\rho
+\epsilon^{\mu\nu\rho}B_\mu\partial_\nu B_\rho-{1\over 4}e F^2
-{e\over 2}(\partial\phi)^2-\alpha_1\,
\epsilon^{\mu\nu\rho}A_\mu\partial_\nu A_\rho  \nonumber\\
&&+ 2\alpha_1\alpha_2e \phi^2 - ie \bar\lambda\gamma^\mu D_\mu \lambda
- 2i \alpha_2 e \bar\lambda\lambda \ \nonumber\\
L_1 &=& e\,\bar\lambda\left[-{i\over 4}\gamma^\mu \gamma^{\nu\rho} F_{\nu\rho}
-{1 \over 2} \gamma^\mu \gamma^\nu \partial_\nu \phi
-\alpha_1 \gamma^\mu \phi \right]\psi_\mu
+ e\, \bar \psi_\mu \left[ -{i \over 4}\gamma^{\nu\rho}  \gamma^\mu F_{\nu\rho}
+ {1 \over 2}  \gamma^\nu \gamma^\mu \partial_\nu \phi
-\alpha_1 \gamma^\mu \phi \right]\lambda \nonumber\\
L_2 &=& -{e \over 2} \bar\psi_\mu \psi_\nu F^{\mu\nu}\phi
-{e \over 4} \bar\lambda\gamma^{\mu\nu}\lambda F_{\mu\nu} \phi
-{ie \alpha_1 \over 2}\bar \psi_\mu \gamma^{\mu\nu}\psi_\nu \phi^2
-{ie\alpha_1 \over 2}\bar\lambda\lambda \phi^2+\alpha_1^2\,e\,\phi^4. \nonumber
\end{eqnarray}
The covariant derivative $D_\mu$ is
$$
D_\mu=\partial_\mu+{1\over 4}\omega_\mu^{ab}\gamma_{ab}
-{1\over 2 l_0}\gamma_\mu-i B_\mu\,;
$$
$F_{\mu\nu}$ is the field strength of $A_\mu$.
The Lagrangian is invariant, up to a total derivative,
under the following linearized supersymmetry transformations,
\begin{eqnarray}
\delta e^a_\mu &=& {i\kappa \over 2} \left(\bar\epsilon \gamma^a \psi_\mu
-\bar\psi_\mu\gamma^a\epsilon\right) \nonumber\\
\delta\psi_\mu &=& {1\over \kappa} D_\mu \epsilon
-\kappa\left({ie\over 4}\epsilon_{\mu\nu\rho}F^{\nu\rho}\phi
+{\alpha_1\over 2}\phi^2\gamma_\mu \right)\epsilon \nonumber\\
\delta B_\mu &=& -{1\over 2\kappa} (\bar\epsilon \psi_\mu
-\bar\psi_\mu\epsilon) \nonumber\\
\delta A_\mu &=& {i\over 2}(\bar\epsilon \gamma_\mu\lambda
- \bar\lambda \gamma_\mu\epsilon )-{\kappa \over 2}
(\bar\epsilon \psi_\mu - \bar\psi_\mu\epsilon)\phi\nonumber\\
\delta \phi &=& {1\over 2}(\bar\epsilon\lambda-\bar\lambda\epsilon)\nonumber\\
\delta \lambda &=&-{1\over 4}F_{\mu\nu}\gamma^{\mu\nu}\epsilon
+{i\over2}\partial_\mu\phi\gamma^\mu\epsilon+i \alpha_1\phi\epsilon \label{dl}
\end{eqnarray}
provided that the parameters $\alpha_1, \alpha_2$ and $l_0$
satisfy the condition
\begin{equation}\label{cond}
\alpha_1 + \alpha_2 = {1 \over l_0}.
\end{equation}
This action thus has two arbitrary parameters, the cosmological radius $l_0$
and the Chern-Simons coupling $\alpha_1$.

For the  construction of the Lagrangian and of the transformation rules
we followed the standard Noether method. $L_{-2}$ is the
gravitational part. $L_0$ contains the kinetic terms and masses of all matter
fields.  $L_1$ are the Noether currents (of global supersymmetry) coupled
to the complex Rarita-Schwinger field. Finally $L_{2}$ ensures linearized
supersymmetry. One may also check that the commutator of two supersymmetry
transformations is a combination of a diffeomorphism
and a gauge transformation.

Setting all fermions and the bosons $B_\mu$ and $\phi$ to zero, the Lagrangian
(\ref{action}) reduces to the bosonic system (\ref{action0}) with $l=l_0$
and $\alpha=\alpha_1$. Thus  the metric and the $U(1)$ gauge field
(\ref{Godel}) and (\ref{gauge}) also solve the equations of motion of the
supersymmetric theory.  However, with this background we meet the same
problems as with the ${\cal N}=1$ theory. With $\phi=0$ we find again
$\delta \lambda \sim F_{\mu\nu} \gamma^{\mu\nu}\epsilon \sim \gamma_0\epsilon
$,
thus, $\delta\lambda=0$ implies $\epsilon=0$.

We observe, however, that the real scalar field in the ${\cal N}=2$
supersymmetric theory has a potential
\begin{equation}\label{V}
V(\phi)=-2\alpha_1\alpha_2\phi^2-\kappa^2\alpha_1^2 \phi^4.
\end{equation}
This means that it could develop a non-zero vacuum expectation value
\begin{equation}\label{vev}
\phi_0=\pm\sqrt{{-\alpha_2\over\kappa^2\alpha_1} }.
\end{equation}
provided that the ratio $\alpha_2/\alpha_1$ is negative ($\phi$ is a real
scalar field). Let's assume this condition holds such that the vev exists.
In this case $\phi_0$ are the two maxima of a potential which is unbounded
from below. $\phi=0$ is a local minimum.

The vev $\phi_0$ has two effects. First, it contributes
non-trivially to the supersymmetry transformations (\ref{dl}),
in particular of the fermionic fields $\lambda$ and $\psi_\mu$.
Second, the value of the potential on $\phi_0$ is not zero.
Setting all fermions to zero and $B_\mu=0$ and $\phi=\phi_0$
we recover the action
(\ref{action0}) with $\alpha=\alpha_1$ and a shifted value for the
cosmological constant:
\begin{eqnarray}\label{1/l}
{1\over l^2} &=& {1\over l_0^2}-\kappa^2V(\phi_0) \nonumber\\
&=& {1 \over l_0^2}-{\alpha_2^2}
\end{eqnarray}

Note that the effective cosmological constant $1/l^2$ can be positive,
negative, or zero. We will see below that in all three cases half of the
supersymmetries are preserved. Of course, there arises the question of
stability of this background. We will not try to answer it here but one
should keep in mind that experience with the AdS vacuum tells us that
the stability properties of fields in non-trivial backgrounds should
be analyzed with care \cite{BF}.  de Sitter supergravity theories in three dimensions have been studied in \cite{LWitten}.

We will now analyze the question whether the background specified by
eqs.(\ref{Godel},\ref{gauge},\ref{vev}), with all other fields set to zero,
preserves some supersymmetry. For convenience we will set
\begin{equation}\label{kappa}
\kappa^2 = {1 \over 2}
\end{equation}
from now on.

The bosonic fields are of course invariant because all fermions are zero
on the background. The variations of $\lambda$ and $\psi_\mu$ give rise
to the equations
\begin{eqnarray}
\delta\lambda=0\qquad&\Rightarrow&
\quad\,\,-{1\over4}F_{\mu\nu}\gamma^{\mu\nu}\epsilon
+{i\over2}\partial_\mu\phi\gamma^\mu\epsilon+i\alpha_1\phi\epsilon = 0
\label{dlambda} \\
\delta\psi_\mu=0\qquad&\Rightarrow& D_\mu \epsilon
-{1\over 2}\left({ie\over4}\epsilon_{\mu\nu\rho}F^{\nu\rho}\phi
+{\alpha_1\over2}\phi^2\gamma_\mu\right)\epsilon = 0 \label{dpsi}
\end{eqnarray}
which must be evaluated on the background defined by eqs.(\ref{Godel}),
(\ref{gauge}), and (\ref{vev}) and, for supersymmetry to be preserved,
must have a nontrivial solution for the supersymmetry parameter $\epsilon$.

We start by evaluating (\ref{dlambda}), which is purely algebraic.
A straightforward calculation gives the condition
\begin{equation}\label{ddel}
(i\alpha_1\phi l-\sqrt{\alpha_1^2 l^2-1}\gamma_0)\epsilon=0\,.
\end{equation}
where $\gamma_0$ is the Dirac matrix with flat (i.e.tangent space) index.
Since $\gamma_0^2=-1$, it has eigenvalues $\pm i$ and we find that
eq.(\ref{ddel}) requires
\begin{equation}
\phi=\pm{1\over\alpha_1 l}\sqrt{\alpha_1^2 l^2-1}
\end{equation}
which agrees with the extrema of the potential, eq.(\ref{vev}).

Next, we analyze the Killing spinor equations (\ref{dpsi}). Using that
$\epsilon$ is an eigenvector of $\gamma_0$ with eigenvalue $\pm i$
we obtain the three equations
\begin{eqnarray}
\partial_t\epsilon&=&\mp{i\over2\alpha_1 l^2}(\alpha_1^2 l^2 +1)\epsilon \\
\partial_r \epsilon &=& 0 \\
\partial_\varphi \epsilon &=& \pm {i \over 2}\epsilon
\end{eqnarray}
which can be easily solved
\begin{equation}
\epsilon=\epsilon_0^\pm
e^{{\pm i\over2}{1+\alpha_1^2 l^2\over\alpha_1 l^2}t\mp{i\over 2}\phi}\,.
\end{equation}
where $\epsilon_0^\pm$ are constant eigenspinors of the flat $\gamma_0$ with
eigenvalues $\pm i$.

The third equation indicates that
$\epsilon(t,\varphi)$ is periodic in $\varphi$ with period $4\pi$, as it
must be for a regular spinor \cite{Henneaux}.

The G\"odel background defined by eqs.(\ref{Godel},\ref{gauge},\ref{vev}) is
thus a supersymmetric solution.  For a given choice of the vev (\ref{vev}),
i.e. for a given sign, there exists one Killing spinor.
In that sense, this solution preserves half of the supersymmetries.

~

Let us now comment on the supersymmetric properties of the other
solutions to the action (\ref{action0}), constructed from (\ref{Godel})
via identifications \cite{BBCG}.  As pointed
out in that reference, the theory described by the action (\ref{action0})
has two sectors $\alpha^2l^2 \geq 1$ and $\alpha^2l^2 <1$.
For $\alpha^2l^2 \geq 1$,
the identifications produce `particles' which have conical singularities
at the origin.  The Killing spinors on this backgrounds do not have the
right periodicity, neither periodic nor antiperiodic, and thus are singular
in the quotient manifold. In other words there are no globally defined
supercharges. More details on this point can be found in \cite{BTZ,CH}.

Identifications on the sector $\alpha^2l^2<1$  produce black holes \cite{BBCG}.
 However, it turns out that the vev (\ref{vev}) for the scalar field is
real only in the sector $\alpha^2l^2 >1$. In fact, from (\ref{1/l}) and
(\ref{cond})
we can express $\alpha_2$ and $l_0$ as functions of $\alpha_1$ and $l$:
\begin{eqnarray}
{1\over l_0} &=& {1 \over 2\alpha_1}\left( {1 \over l^2} + \alpha_1^2 \right) \\
\alpha_2 &=& {1\over2\alpha_1}\left({1\over l^2}-\alpha_1^2\right)\label{alpha2}
\end{eqnarray}
From  (\ref{alpha2}) we conclude that
$-2\alpha_2/\alpha_1 = (\alpha_1^2l^2 -1)/\alpha_1^2l^2$.
Thus, the vev (\ref{vev}) is real only in the sector $\alpha_1^2l^2 >1$.

It would be interesting to find a supergravity theory yielding
supersymmetric backgrounds for $\alpha^2l^2 <1$.  In that sector black
holes are present and one could then ask whether extreme ones are
supersymmetric or not.

We conclude with some comments. The action we have constructed is
supersymmetric at the linear order. In principle, the higher fermionic terms in
the action and the transformation rules can be constructed via the Noether
procedure. But this is very tedious. A more promising approach is to use
superfields and we leave this for the future. Another immediate question
is how the three dimensional G\"odel background which we have studied here
can be obtained from the five-dimensional solution of \cite{Gauntlett-}
via compactification.
Finally, as we have already
mentioned, the stability issue might be worth studying.

\bigskip

\noindent
{\bf Acknowledgments}

MB would like to thank H. Nicolai for hospitality at the Max Planck
Institute (Potsdam) where part of this work was done. MB was also
partially supported by grants Fondecyt (Chile) \#1060648 and \#1051084.

\end{document}